\documentclass[12pt,english]{article}
\usepackage[T1]{fontenc}
\usepackage[koi8-r]{inputenc}
\usepackage{amsmath}
\usepackage{esint}

\makeatletter
\@ifundefined{date}{}{\date{}}

\newtheorem{theorem}{Theorem}

\newtheorem{corollary}{Corollary}

\newtheorem{lemma}{Lemma}

\usepackage{babel}

\makeatother

\usepackage{babel}
\begin{document}

\title{Self-interaction model of classical point particle in one-dimension }

\author{Malyshev V. A.%
\thanks{Faculty of Mechanics and Mathematics, Lomonosov Moscow State University.
Vorobievy Gory, Main Building, 119991, Moscow Russia, 2malyshev@mail.ru%
} , Pirogov S. A.%
\thanks{Institute of information Transmission Problems, Bolshoj Karetnyj,
19, Moscow, Russia, s.a.pirogov@bk.ru%
}}
\maketitle
\begin{abstract}
We consider a hamiltonian system on the real line, consisting of real
scalar field $\phi(x,t)$ and point particle with trajectory $y(t)$.
The dynamics of this system is defined by the system of two equations:
wave equation for the field, <<radiated>> by the point particle,
and Newton's equation for the particle in its own field. We find the
solution where the particle is strongly damped, but the kinetic and
interaction energies of the field increase linearly in time, in despite
of the full energy conservation..
\end{abstract}

\paragraph{Introduction}

In classical physics the matter (for example, point particles) and
continuous fields are linked with two types of equations: 1) fields
move the particles, 2) particles generate fields. The first type are
the Newton equations, a particular case is the Lorentz equation (Lorentz
force). The second are based on the Maxwell equations with fixed trajectories
of the point charges. The problem of joining these two systems together
always, starting possibly with \cite{Lamb}. drew much attention of
physicists, but still rests terra incognita. Possible approaches to
this problem can differ globally and in small details. For example,
one can introduce additional forces, keeping smeared charge inside
balls, as in the Abraham model (see bibliography in the book \cite{Spohn}).
We want also to mention another mathematical model of particle-field
interaction \cite{Komech}, however it is sufficiently far away from
our model. 

In this paper we study seemingly the simplest self-interaction model
without introducing additional forces. The main interest of this model
is not only that it is hamiltonian, allows rigorous analysis and appears
to be almost explicitely solvable, but rather that the energy of the
field and of the interaction energy, taken separately, are not bounded
(in despite of the energy conservation). This fact seems to be important
but demands further comprehension.

Note that the considered model of particle-field interaction is a
non-relativistic analog of the scalar gravity theory by G. Nordstrom
\cite{scalar}.

\paragraph{The model}

We consider a system on the real line, consisting of the real scalar
field $\phi(x,t),x\in R,t\in R_{+},$ and point particle with trajectory
$y(t)\in R$. The dynamics of this system is defined by two equations:
wave equation for the field, <<radiated>> by the particle,

\begin{equation}
\frac{\partial^{2}\phi(x,t)}{\partial t^{2}}=c^{2}\frac{\partial^{2}\phi(x,t)}{\partial x^{2}}+\beta f(x-y(t))\label{eq_field_general}
\end{equation}
with the initial conditions 
\begin{equation}
\phi(x,0)=0,\phi'_{t}(x,0)=0,\label{initial_field}
\end{equation}
and the Newton equation for the particle, driven by its own field,
\begin{equation}
m\frac{d^{2}y(t)}{dt^{2}}=\beta\frac{\partial}{\partial y}\int_{-\infty}^{\infty}\phi(x,t)f(x-y)dx\label{eq_particle_general}
\end{equation}
with the initial conditions 
\begin{equation}
y(0)=0,\frac{dy}{dt}(0)=v(0)=v_{0}\label{initial_particle}
\end{equation}

It is well known, see for example \cite{Vladimirov}, that for any
locally integrable $f$, and given smooth $y(t)$ the unique solution
$\phi(x,t)$ of the linear inhomogeneous equation (\ref{eq_field_general})
with initial conditions (\ref{initial_field}) is also locally integrable
and can be written as 
\begin{equation}
\phi(x,t)=\frac{\beta}{2c}\int_{0}^{t}\int_{x-c(t-\tau)}^{x+c(t-\tau)}f(x_{1}-y(\tau))dx_{1}d\tau\label{main_solution}
\end{equation}
However the joint system of these equations is nonlinear, and we do
not know general rigorous results concerning the structure of its
solutions. 

\begin{lemma}\label{lemma_smooth}

If the function $f$ is smooth and bounded, then the solution of the
system (\ref{eq_field_general})-(\ref{initial_particle}) exists
and is unique on all time interval $[0,\infty)$.

\end{lemma}

The goal of this paper is to give exact sense and get complete picture
of the dynamics for the ultra-local interaction, that is for the case
when $f$ is the $\delta$-function. 

\begin{theorem}

If $f=\delta$ and $|v_{0}|<c$, then there exists a solution $(\phi(x,t),y(t))$
of the eqautions (\ref{eq_field_general})-(\ref{initial_particle})
in the domain $x\in R,t\in[0.\infty),$ such that $v(t)=y'(t)$ is
a smooth monotone function on $[0,\infty)$. For this solution 
\begin{equation}
\sup_{0\leq t<\infty}|v(t)|<c\label{sup_condition}
\end{equation}
and $v(t)\to0$ exponentially fast as $t\to\infty$. Moreover, this
solution is unique in the class of smooth solutions, satisfying condition
(\ref{sup_condition}).

\end{theorem}

\paragraph{Energy}

The equations (\ref{eq_field_general}) and (\ref{eq_particle_general})
can be written in the hamiltonian form
\begin{equation}
\frac{\partial^{2}\phi(x,t)}{\partial t^{2}}=-\frac{\delta U}{\delta\phi(x)},\,\,\, m\frac{d^{2}y(t)}{dt^{2}}=-\frac{\partial U}{\partial y},U=U_{ff}+U_{fp}\label{U_energy}
\end{equation}
with the formal hamiltonian $H=T_{f}+T_{p}+U_{ff}+U_{fp}$, where
\begin{equation}
T_{f}=\int\frac{1}{2}(\frac{\partial\phi}{\partial t})^{2}dx,\,\,\, T_{p}=\frac{m}{2}v^{2}\label{T_energy}
\end{equation}
are the kinetic energies of the field and of the particle, and 
\[
U_{ff}=\int[\frac{c^{2}}{2}(\frac{\partial\phi}{\partial x})^{2}dx,\,\,\, U_{fp}=-\beta\int\phi(x)f(x-y)]dx=-\beta\phi(y)
\]
where $U_{ff}$ is the self-interaction energy of the field, $U_{fp}$
is the particle-field interaction. 

\begin{theorem}

Let $f=\delta$. Then for any fixed $t$, the supports of the derivatives
$\frac{\partial\phi}{\partial x},\frac{\partial\phi}{\partial t}$
are bounded in $x$, and all energy constituents are finite and have
the following asymptotics as $t\to\infty$ 
\[
T_{p}(t)\to0,\,\,\, T_{f}(t)\sim\frac{\beta^{2}}{4c},\,\,\, U_{ff}(t)\sim\frac{\beta^{2}}{4c}t,
\]
\[
U_{fp}(t)=-\frac{\beta^{2}}{2c}t
\]

\end{theorem}

The energy conservation is proved by the standard calculation
\[
\frac{dH}{dt}=\int[c^{2}\frac{\partial\phi}{\partial x}\frac{d}{dt}(\frac{\partial\phi}{\partial x})+\frac{\partial\phi}{\partial t}\frac{d}{dt}(\frac{\partial\phi}{\partial t})-\beta\frac{\partial\phi}{\partial t}f(x-y(t))-\beta\phi(x,t)\frac{\partial f}{\partial y}v]dx+mv\frac{dv}{dt}=
\]
\[
=\int[-c^{2}\frac{\partial^{2}\phi}{\partial x^{2}}\frac{\partial\phi}{\partial t}+\frac{\partial\phi}{\partial t}\frac{\partial^{2}\phi}{\partial t^{2}}-\beta\frac{\partial\phi}{\partial t}f(x-y(t))-\beta v\phi(x,t)\frac{\partial f(x-y)}{\partial y}+\beta v\phi(x,t)\frac{\partial f(x-y)}{\partial y}]dx=0
\]
where the last two terms mutually cancel, and the first three terms
give zero due to equation (\ref{eq_field_general}).

\paragraph{Proof of theorem 1}

The plan of the proof is the following. By explicit formula (\ref{main_solution}),
one can forget about equation (\ref{eq_field_general}) and, substituting
(\ref{main_solution}) to (\ref{eq_particle_general}), solve the
obtained integro-differential equation. It is not clear how to calculate
integrals for arbitrary $y(t)$, but if one assumes in advance the
condition (\ref{sup_condition}) on $y(t)$, then one gets the solution
of equations (\ref{eq_field_general})-(\ref{initial_particle}),
which miraculously appears to satisfy this assumption. 

Proof of lemma \ref{lemma_smooth}. Local (in time) existence and
uniqueness of the solution can be proved in the standard way. To prove
that the solution exists on all time interval one needs more accurate
estimates. From (\ref{main_solution}) we have 
\begin{equation}
\frac{\partial}{\partial x}\phi(x,t)=\frac{\beta}{2c}\int_{0}^{t}[f(x+c(t-\tau)-y(\tau))-f(x-c(t-\tau)-y(\tau))]d\tau\label{x_derivative}
\end{equation}
From (\ref{main_solution}) it is clear that $|\frac{d\phi}{dx}|$
for given $t$ and all $x$ does not exceed $Bt$, where 
\[
B=\frac{\beta}{c}\sup|f|
\]
This means that the absolute value of the particle acceleration does
not exceed $\frac{B}{m}t$, that is $|y(t)|$ does not exceed $const\, t^{3}$.
Thus there cannot be vertical asymptote for finite $t$. Lemma is
proved.

\begin{lemma}\label{lemma_derivative} Let $y(t)$ be sufficiently
smooth and let the condition (\ref{sup_condition}) be satisfied.
Then 
\begin{equation}
\frac{\partial\phi(x,t)}{\partial x}=\begin{cases}
0, & x\notin[-ct,ct]\\
\frac{\beta}{2c}\frac{1}{c+y'(\tau(x))} & x\in[-ct,y(t))\\
-\frac{\beta}{2c}\frac{1}{c-y'(\tau(x))}, & x\in(y(t),ct]\\
-\frac{\beta}{2c}\frac{y'(t)}{c^{2}-(y'(t))^{2}} & x=y(t)
\end{cases}\label{proizvodnaia_po_x}
\end{equation}

\end{lemma}

Proof. Instead of directly using substituion techniques for $\delta$-function
(as for example in \cite{Misra}), it is more convenient to use the
gaussian approximation for the $\delta$-function 
\[
f_{\sigma}(x,t)=\delta_{\sigma}(x-y(t)),\,\,\delta_{\sigma}(x)=\frac{1}{\sigma\sqrt{2\pi}}e^{-\frac{x^{2}}{2\sigma^{2}}}\to_{\sigma\to0}\delta(x)
\]
In our case

\begin{equation}
\frac{\partial\phi_{\sigma}(x,t)}{\partial x}=\frac{\beta}{2c\sigma\sqrt{2\pi}}\int_{0}^{t}(e^{\frac{h_{+}}{\sigma^{2}}}-e^{\frac{h_{-}}{\sigma^{2}}})d\tau\label{sum_of_integrals}
\end{equation}
where for given $t$
\[
h_{\pm}=h_{\pm}(x,\tau)=-\frac{1}{2}(x-y(\tau)\pm c(t-\tau))^{2}
\]
For given $t$ define the functions $x_{\pm}(\tau)=y(\tau)\mp c(t-\tau)$,
In other words we choose them so that
\[
h_{\pm}(x_{\pm}(\tau),\tau)=0
\]
When $\tau$ runs along the interval $[0,t]$, $x_{\pm}(\tau)$ runs
inside the interval $[\mp ct,y(t)]$ correspondingly. Define also
the functions $\tau_{\pm}(x)$ on the intervals $[-ct,y(t)]$ and
$[y(t),ct]$ correspondingly by the condition
\[
h_{\pm}(x,\tau_{\pm}(x))=0
\]
These functions are also uniquely defined because the functions $y(\tau)\pm c(t-\tau)$
are strictly monotone by our assumption. These functions coincide
at the point $y(t)$ where they equal to $t$. Thus they can be glued
in one function on the interval $[-ct,ct]$, we denote this function
$\tau(x)$. Outside this interval we put $\tau(x)=0$.

Then
\[
h'_{\pm}=\frac{\partial h_{\pm}}{\partial\tau}(x,\tau)=-(x-y(\tau)\pm c(t-\tau))(\mp c-y'(\tau)),
\]

\[
h_{\pm}''=\frac{\partial^{2}h_{\pm}}{\partial\tau^{2}}(x,\tau)=(x-y(\tau)\pm c(t-\tau))y''(\tau))-(\mp c-y'(\tau))^{2}
\]
and at the point $(x,\tau_{\pm}(x))$ we have
\[
\frac{\partial h_{\pm}}{\partial\tau}(x,\tau_{\pm}(x))=-(x-y(\tau_{\pm}(x))\pm c(t-\tau_{\pm}(x)))(\mp c-y'(\tau_{\pm}(x)))=0,
\]

\[
h''(x,\tau_{\pm}(x))=\frac{\partial^{2}h_{\pm}}{\partial\tau^{2}}(x,\tau_{\pm}(x))=-(\mp c-y'(\tau_{\pm}(x)))^{2}
\]
Note that if $x\neq x_{\pm}(\tau)$ for any $\tau\in[0,t]$, then
\[
\int_{0}^{t}e^{\frac{h_{\pm}}{\sigma^{2}}}d\tau\to_{\sigma\to0}0
\]
In particular, this takes place for any $x\notin[-ct,ct]$. Moreover,
$x\neq x_{+}(\tau)$ for all $\tau\in[0,t]$, if $x\notin[-ct,y(t)]$.
Similarly, $x\neq x_{-}(\tau)$ for all $\tau\in[0,t]$, if $x\notin[y(t),ct]$.
For other $x$ we will apply to any integral (\ref{sum_of_integrals})
the Laplace method, using the following result 
\begin{equation}
\int_{a}^{b}e^{nh(\tau)}d\tau\sim e^{nh(a)}[-\frac{\pi}{2nh''(a)}]^{\frac{1}{2}}\label{laplace_2}
\end{equation}
if $h(\tau)$ has its maximum at the point $a$, $h'(a)=0$ and $h''(a)<0$
(similarly for $b$). If the maximum is reached at some point $u$,
lying inside the interval $(a,b)$, and also $h'(u)=0$ and $h''(u)<0$,
then the right hand side in (\ref{laplace_2}) is multiplied by 2. 

In our case $h'(x,\tau_{\pm}(x))=0,h''(x,\tau_{\pm}(x))\neq0$, and
\[
\frac{\beta}{2c\sigma\sqrt{2\pi}}\int_{0}^{t}e^{\frac{h_{+}}{\sigma^{2}}}d\tau\sim\frac{\beta}{c\sigma\sqrt{2\pi}}\sigma(\frac{\pi}{2})^{\frac{1}{2}}\frac{1}{c+y'(\tau(x))},x\in[-ct,y(t)]
\]
\[
\frac{\beta}{2c\sigma\sqrt{2\pi}}\int_{0}^{t}e^{\frac{h_{-}}{\sigma^{2}}}d\tau\sim\frac{\beta}{c\sigma\sqrt{2\pi}}\sigma(\frac{\pi}{2})^{\frac{1}{2}}\frac{1}{c-y'(\tau(x))},x\in[y(t),ct]
\]
and for $x=y(t)$ as $\sigma\to0$ 
\[
\frac{\partial\phi_{\sigma}(y(t),t)}{\partial x}\to\frac{\beta}{2c\sigma\sqrt{2\pi}}\sigma(\frac{\pi}{2})^{\frac{1}{2}}(\frac{1}{c+y'(t)}-\frac{1}{c-y'(t)})=-\frac{\beta}{2c}\frac{y'(t)}{c^{2}-(y'(t))^{2}}
\]
Lemma is proved.

Let us prove now the first assertion of the theorem 1. By lemma \ref{lemma_derivative},
the equation (\ref{eq_particle_general}) for the particle becomes
\[
m\frac{d^{2}y(t)}{dt^{2}}=\beta\frac{\partial}{\partial y}\phi(y(t),t)=-\frac{\beta^{2}}{2c}\frac{y'(t)}{c^{2}-(y'(t))^{2}}
\]
or 
\begin{equation}
m\frac{dv}{dt}=-\frac{\beta^{2}}{2c}\frac{v}{c^{2}-v^{2}}\label{v_equation}
\end{equation}
This means that $v(t)$ tends to the fixed point $v=0$ exponentially
fast.

\begin{corollary}

The solution $(\phi_{\sigma}(x,t),f_{\sigma}(t))$, which accordingly
to \ref{lemma_smooth}, exists, unique, and converges (as $\sigma\to0$)
to the solution obtained in theorem 1.

\end{corollary}

\paragraph{Proof of theorem 2}

If $f_{\sigma}\to\delta$, then
\[
\phi_{\sigma}(x,t)\to\frac{\beta}{2c}\int_{0}^{t}\mathbf{1}\{x:y(\tau)-c(t-\tau)<x<c(t-\tau)+y(\tau)\}d\tau=\frac{\beta}{2c}\tau(x)
\]
and, by $\tau(y(t))=t$, we get
\[
U_{fp}=-\beta\phi(y(t),t)=-\frac{\beta^{2}}{2c}t
\]
Further on, using lemma \ref{lemma_derivative} and taking into account
that $v(t)\to0$ as $t\to\infty$, we have 
\[
U_{ff}=\frac{c^{2}}{2}\int(\frac{\partial\phi}{\partial x})^{2}dx=\frac{c^{2}}{2}[\int_{-ct}^{y(t)}(\frac{\beta}{2c}\frac{1}{c-y'(\tau(x))})^{2}dx+\int_{y(t)}^{ct}(\frac{\beta}{2c}\frac{1}{c+y'(\tau(x))})^{2}dx]\sim\frac{\beta^{2}}{4c}t
\]
Finally
\begin{equation}
\frac{\partial}{\partial t}\phi_{\sigma}(x,t)=\frac{\beta}{2}\int_{0}^{t}(f(x+c(t-\tau)-y(\tau))+f(x-c(t-\tau)-y(\tau)))d\tau\label{proiz_po_vremeni}
\end{equation}
and in the limit $f=\delta$ 
\[
\frac{\partial}{\partial t}\phi_{\sigma}(x,t)=c\frac{\partial\phi_{\sigma}(x,t)}{\partial x}sign(y(t)-x)
\]
That is why

\[
T_{f}=U_{ff}\sim\frac{\beta^{2}}{4c}t,
\]
that also follows (as we know asymptotics of other energy constituents)
from the energy conservation.

\end{document}